\documentstyle[twocolumn,epsf]{jpsj}

\def\rv{\mbox{\boldmath $r$}}
\def\xv{\mbox{\boldmath $x$}}
\def\ep{\epsilon}

\def\runtitle{Letters}
\def\runauthor{Letters}

\title
{
Josephson Current Flowing in Cyclically Coupled
Bose-Einstein Condensates
}

\author
{
Makoto {\sc Tsubota}
and Kenichi {\sc Kasamatsu} 
}

\inst
{
Department of Physics, Osaka City University, Sumiyoshi 558-8585\\
}

\recdate
{
\today
}

\abst
{
The Josephson effect in cyclically coupled Bose-Einstein condensates
is studied theoretically. We analyze the simultaneous Gross-Pitaevskii
equations with coupling terms between adjacent condensates.
Depending on the initial relative phases between condensates, Josephson
current flows cyclically to create a quantized vortex.
Reducing the coupling between condensates changes the motion from periodic
to chaotic, thus suppressing the cyclic current.
The relation to the Kibble-Zurek mechanism is discussed; the density of 
the generated vortices is less than that predicted by this mechanism.
}

\kword
{
Josephson effect, Kibble-Zurek mechanism, vortex generation,
Bose-Einstein condensation
}

\begin{document}
\setlength{\evensidemargin}{-10pt}
\setlength{\oddsidemargin}{-10pt}
\sloppy
\maketitle

The experimental observation of the Bose-Einstein condensation
 (BEC) in a dilute gas of neutral, trapped atoms opens up 
studies of interesting quantum phenomena on the macroscopic scale
\cite{Review}. The broken gauge symmetry due to BEC produces a
macroscopic wave function (the order parameter) $\Psi(\rv,t)=
\sqrt{\rho}e^{\imath \theta}$, with $\rho$ as the condensate density
and $\theta$ as its phase.
This wave function obeys the Gross-Pitaevskii equation (GPE), which
has successfully explained many phenomena of the BEC of neutral
atoms.
One of the significant effects caused by this wave function is the 
Josephson effect.
Raghavan et al. theoretically studied the coherent atomic tunneling
between two weakly coupled BECs\cite{Rag}.
They analyzed the Feynman two-state model derived from the GPE to
investigate the characteristic nonlinear phenomena which are not
accessible with superconductor Josephson junctions.

The first motive of this work is to improve the density and phase dynamics
for multicoupled BECs; this theoretical study can lead to new experiments.
The second motive is closely related to the experimental observation of
vortex
generation in superfluid $^3$He and $^4$He.
Symmetry-breaking phase transitions in the early universe immediately 
after the Big Bang are expected
to leave behind long-lived topologically stable structures such as monopoles,
strings or domain walls.
Zurek discussed the analogy of the second-order phase transition between
cosmological strings and quantized vortices in the superfluid,
and suggested a cryogenic experiment which tested the cosmological scenario
for string formation\cite{Zurek}.
The Kibble-Zurek (KZ) scenario is the following: after the system enters the
broken
symmetry phase through a rapid phase transition, the ordered phases 
start to form, simultaneously and independently, in many parts of the system. 
Each ordered region has an independent phase $\theta$ of the
order parameter field $\Psi(\rv,t)$, because they are disconnected causally 
from each other. Critical slowing down near the transion temperature 
suppresses the velocity with which the order parameter coherence spreads.
Consequently, when the disconnected domains grow together and superfluid 
coherence becomes established throughout the entire volume, the topological 
defects around which supercurrent flows, i.e., the vortices, are frozen 
out at the boundaries. Zurek estimated the defect density from the coherent 
length which is comparable to the frozen domain size. it should be noted 
that this estimation is based on the assumption that the boundary between 
more than three domains always creates a defect.
This theoretical scenario motivated various experimental works.
The experiments\cite{Bau,Ruu} which heated small regions in superfluid
$^3$He by a
neutron-induced nuclear reaction yielded results consistent with the
theoretical
estimation of the vortex tangle density, while Volovik subseqently noted 
that the experimental work\cite{Ruu} using rotating superfluid $^{3}$He 
evidenced not only the vortices generated by the KZ mechanism but also 
those induced by the rotating flow\cite{Vol}. The experiment\cite{Mc} which
expanded 
liquid $^4$He rapidly through the lambda line observed a density that was at least
two orders of magnitude less than the theoretical estimation;
the reason for the disagreement remains unresolved. Hence, at present, 
the KZ scenario is not necessarily confirmed experimentally.

Thus, this paper discusses one important factor which may
be omitted in the above scenario. A cyclic supercurrent should flow through 
adjacent weakly coupled domains in order for a vortex to appear there. 
However, it is not clear whether
cyclically coupled
domains necessarily result in cyclic Josephson current (CJC) or not.
Therefore, by improving the physics of two weakly coupled BECs, this work
studies
three cyclically coupled BECs. Whether CJC appears or not is found to depend
on the initial relative phases and the coupling constants between
adjacent BECs. It is impossible to control the initial relative phases.
Hence, if initial relative phases are assumed to take arbitrary 
values between 0 and $2\pi$
with equal probability, we obtain the expected value of CJC which depends
on the coupling constants. By applying these results
to more than three cyclically
coupled BECs, we find that the density of generated vortices may be reduced
compared with the original KZ scenario.

First, the formulation for the two weakly coupled BECs is described
briefly\cite{Rag}.
The macroscopic wave function $\Psi(\rv,t)$ in a trap potential $V(\rv)$
at $T=0$ satisfies the GPE
\begin{eqnarray}
\imath \hbar \frac{\partial \Psi(\rv,t)}{\partial t}&=&-\frac{\hbar^2}{2m}
\nabla^2 \Psi(\rv,t) \nonumber\\
&&+[V(\rv)+g_0 |\Psi(\rv,t)|^2]\Psi(\rv,t),
\end{eqnarray}
with $g_0=4\pi \hbar^2 a/m$, $m$ the atomic mass, and $a$ the $s$-wave
scattering length of the atoms.
When $V(\rv)$ is a double-well trap, $\Psi(\rv,t)$ is written as
\begin{equation}
\Psi(\rv,t)= \psi_1(t)\Phi_1(\rv)+\psi_2(t)\Phi_2(\rv),
\end{equation}
where $\Phi_{1,2}$ describe the condensate which is localized in each trap,
$\psi_{1,2}(t)=\sqrt{N_{1,2}}e^{\imath\theta_{1,2}(t)}$ and
the total number of atoms $N_T=N_1+N_2=|\psi_1|^2+|\psi_2|^2$.
The weak coupling causes the overlap integral $\int \Phi_1(\rv)
\Phi_2(\rv)d\rv$ to be negligible, which leads to the nonlinear two-mode
dynamical equations
\begin{subequations}
\begin{eqnarray}
\imath \hbar \frac{\partial \psi_1}{\partial t}=(E_1^0+U_1N_1)\psi_1
-K\psi_2 \\
\imath \hbar \frac{\partial \psi_2}{\partial t}=(E_2^0+U_2N_2)\psi_2
-K\psi_1.
\end{eqnarray}
\end{subequations}
Here, $E_{1,2}^0$ are the zero-point energies in each well, $U_{1,2}N_{1,2}$
the atomic self-interaction energies and $K$ the amplitude of the tunneling
between condensates.
Introducing the population imbalance $z(t) \equiv (N_1(t)-N_2(t))/N_T$ and
the relative phase $\phi(t) \equiv \theta_2(t)-\theta_1(t)$, and rescaling
to a dimensionless time $2Kt/\hbar \rightarrow t$, eqs.(3a) and (3b) are 
reduced to the
canonical form:
\begin{equation}
\frac{dz}{dt}=-\frac{\partial H}{\partial \phi}, \hspace{8mm}
\frac{d\phi}{dt}=\frac{\partial H}{\partial z}.
\label{Hami}
\end{equation}
Here, the Hamiltonian is
\begin{equation}
H=\Delta E z +\frac{\Lambda z^2}{2}-\sqrt{1-z^2}\cos \phi,
\end{equation}
and the parameters are
\begin{equation}
\Lambda=\frac{(U_1+U_2)N_T}{4K},\hspace{1mm}
\Delta E = \frac{E_1^0-E_2^0}{2K}+\frac{(U_1-U_2)N_T}{4K}.
\end{equation}
Although the exact solutions of eq.\ (\ref{Hami}) are studied in detail in
Ref.2, investigating the conserved energy $H$ as a function
of $z(t)$ and $\phi(t)$ qualitatively reveals its dynamics.
One characteristic motion is "macroscopic quantum self-trapping" (MQST); the
average population difference between two BECs is nonzero.
Another is $\pi$-phase oscillations which are observed experimentally
in the weak link separating superfluid $^3$He-B reservoirs\cite{Packard}.
These Josephson $\pi$ states and their quantum decay have also been studied 
by Hatakenaka\cite{Hata}.

Extending the above formulation, we will study $n$ cyclically coupled BECs.
The substitution of the representation $\Psi(\rv,t)=\sum_i \psi_i(t)
\Phi_i(\rv)$
into eq. (1) yields
\begin{eqnarray}
\imath \hbar \frac{\partial \psi_i}{\partial t}=(E_i+U_iN_i)\psi_i
-K_{i,i-1}\psi_{i-1}-K_{i,i+1}\psi_{i+1}, \nonumber\\
\hspace{40mm}(i=1,2,\cdots,n)
\label{neq}
\end{eqnarray}
where the cyclic boundary condition means $\psi_0 = \psi_n$ and
$\psi_{n+1} = \psi_1$. The parameter $K_{i,i+1}$ is the coupling constant
between $i$th and $i+1$th BEC.
The simplest case of $n=3$ is discussed. The insertion of
$\psi_i=\sqrt{N_i}e^{\imath \theta}$ in eq. \ (\ref{neq}) yields
\begin{subequations}
\begin{eqnarray}
\hbar\dot{N_1} & = &2K_{12}\sqrt{N_1N_2}\sin \phi_{12}
-2K_{31}\sqrt{N_3N_1}\sin \phi_{31} \\
\hbar\dot{N_2} & = & 2K_{23}\sqrt{N_2N_3}\sin \phi_{23}
-2K_{12}\sqrt{N_1N_2}\sin \phi_{12} \\
\hbar\dot{N_3} & = & 2K_{31}\sqrt{N_3N_1}\sin \phi_{31}
-2K_{23}\sqrt{N_2N_3}\sin \phi_{23} \\
\hbar\dot{\theta_1} & = & - E_1-U_1N_1  \nonumber\\
& & +K_{12}\sqrt{\frac{N_2}{N_1}}\cos \phi_{12}
+K_{31}\sqrt{\frac{N_3}{N_1}}\cos \phi_{31} \\
\hbar\dot{\theta_2} & = & - E_2-U_2N_2  \nonumber\\
& & +K_{23}\sqrt{\frac{N_3}{N_2}}\cos \phi_{23}
+K_{12}\sqrt{\frac{N_1}{N_2}}\cos \phi_{12} \\
\hbar\dot{\theta_3} & = & - E_3-U_3N_3  \nonumber\\
& & +K_{31}\sqrt{\frac{N_1}{N_3}}\cos \phi_{31}
+K_{23}\sqrt{\frac{N_2}{N_3}}\cos \phi_{23},
\end{eqnarray}
\label{equ}
\end{subequations}
where $\phi_{ij}=\theta_i-\theta_j$ is a relative phase.
The Josephson supercurrent which flows from the $i$th 
to the $j$th condensate is expressed as
\begin{equation}
J_{i \rightarrow j}=-\frac{2K_{ij}}{\hbar}\sqrt{N_i N_j}\sin \phi_{ij}.
\end{equation}
Equations (8a)-(8f) conserve the total number
$N_T=N_1+N_2+N_3$ and the total energy
\begin{equation}
E=\sum_i (E_iN_i+\frac12 U_i N_i^2)+\sum_{i<j} 2K_{ij} \sqrt{N_iN_j} \cos
\phi_{ij}.
\end{equation}
Since the wave function $\Psi(\rv,t)$ is single-valued,
the relative phases must satisfy
\begin{equation}
\oint_{\cal C} \nabla \phi d\mbox{\boldmath $\ell$}
=\phi_{12}+\phi_{23}+\phi_{31}=2\pi m,
\label{phase}
\end{equation}
where $\cal C$ is a contour that passes cyclically through the
condensates and $m$ an integer.
The variable $\phi_{31}$ is eliminated by eq.\ (\ref{phase}).
Equations (8a)-(8f) show that
the dynamics is independent of $m$.
The following calculations are performed for the dimensionless
variables:
\begin{equation}
\frac{Kt}{\hbar}\rightarrow t, \hspace{5mm} \frac{N_i}{N_T}
\rightarrow n_i, \hspace{5mm} \frac{\hbar J_{i\rightarrow j}}
{KN_T} \rightarrow j_{i\rightarrow j}.
\end{equation}

We will study the dynamics for $E_1=E_2=E_3=E'$ and $K_1=
K_2=K_3=K$.  Then, the parameters for the dimensionless equations
are $\ep=E'/K$, and $\Lambda_i=U_iN_T/K$ that represents the
strength of the coupling.  We assume
$\Lambda_1=\Lambda_2=\Lambda_3=\Lambda$.
Attention will be paid mainly to whether CJC flows or not.
First, under $\Lambda=1$ and the fixed initial values
$n_1(0)=n_2(0)=n_3(0) = 1/3$, the initial phases $\phi_{12}(0)$ and
$\phi_{23}(0)$ are changed.
The oscillation of $j_{i\rightarrow j}$ is shown in Fig. 1.
In Fig. 1(a) for $\phi_{12}(0)=\phi_{23}(0)=\pi/8$,
each current oscillates about zero; net currents
flow neither clockwise nor counterclockwise.
However, when the initial phases are changed (Figs. 1(b) and 1(c)),
the time-averaged value of every current becomes negative,
which means that CJC appears.

\begin{figure}
\begin{center}
\epsfxsize=9cm \ \epsfbox{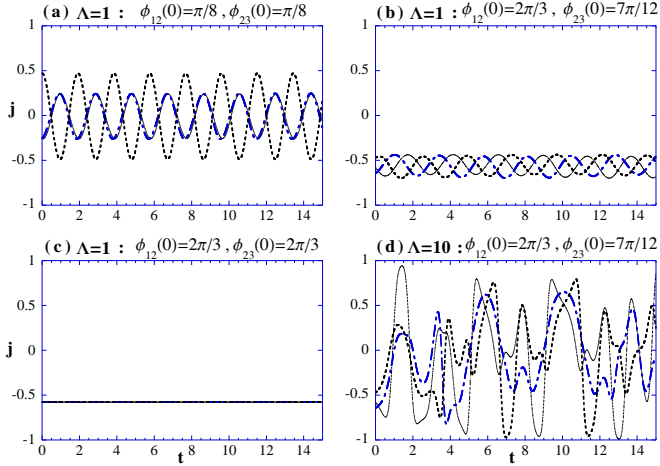} \\
\vspace{3mm}
\caption{Oscillation of $j_{1\rightarrow 2}$ (solid line),
$j_{2\rightarrow 3}$ (dashed-dotted line) and $j_{3\rightarrow 1}$ (dotted line).
Values of $\Lambda$ and initial relative
phases are shown in each figure.}
\end{center}
\end{figure}

The appearance of CJC is found by investigating
the time- and piece-averaged current
\begin{equation}
<j>_{av}= \frac{1}{3T} \int_0^T dt (j_{1\rightarrow 2}
+j_{2\rightarrow 3}+j_{3\rightarrow 1}),
\label{CJC}
\end{equation}
where $T$ is a period of time which is sufficiently longer than the
period of oscillation.

Figure 2(a) shows how $<j>_{av}$ depends on $\phi_{12}(0)$ and
$\phi_{23}(0)$ at $\Lambda =1$.
The current $<j>_{av}$ has two extremums for $\phi_{12}(0)=\phi_{23}
(0)=2\pi/3$ and $4\pi/3$ which give the stationary solutions.
However, when $\Lambda$ is increased, i.e., the coupling is reduced,
as shown in Figs. 1(b) and 1(d), the oscillation of currents become chaotic;
the transition from periodic to chaotic motion 
is confirmed by the analysis of the power spectrum.
The important property of the $n\geq 3$ BECs dynamics is this
transition to chaos which is absent in two coupled BECs\cite{Rag}.
When the motion becomes chaotic, $<j>_{av}$ is lessened considerably.
Thus, it is concluded that the appearance of
 CJC depends on the initial relative
phases and the coupling constants, which are summarized in Fig. 2.
Since we cannot control the initial relative phases, CJC appears
with some probability that depends on the coupling constants.
If every relative phase is assumed to take a value between 0 and
$2\pi$ with equal probability, the expected values or the ensemble averages
of $<j>_{av}$
for $\Lambda=$1, 3 and 10 are, respectively, 0.174, 0.088 and 0.029.
This system has some unique modes which are compared with those of
two coupled BECs\cite{Rag}.
The oscillative ones are zero-phase, $2\pi/3$-phase and $4\pi/3$-phase
modes in which the time-average of the phase $<\phi_{ij}>=0$, $2\pi/3$
and $4\pi/3$, respectively.
There are also running-phase modes that create MQST.
The analysis of the linear stability finds that the zero-phase mode
is stable for an arbitrary value of $\Lambda$, while the $2\pi/3$- and
$4\pi/3$-phase modes become unstable for $\Lambda > 9/4$.

\begin{figure}
\begin{center}
\epsfxsize=7.5cm \ \epsfbox{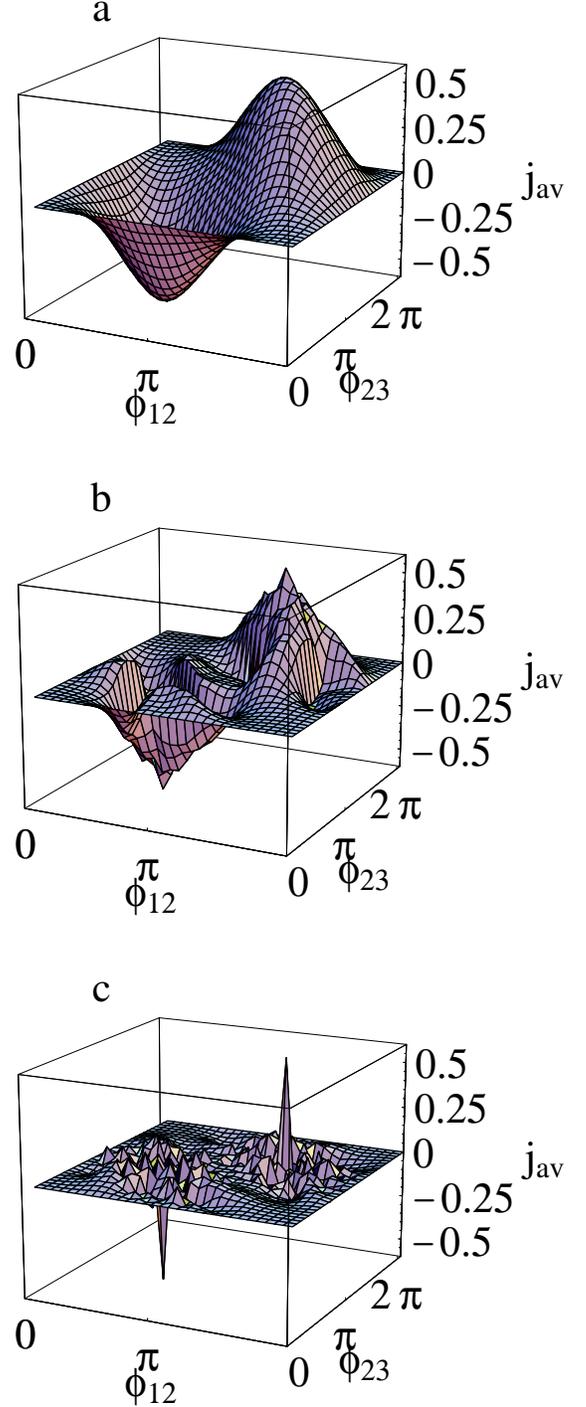} \\
\vspace{3mm}
\caption{Dependence of the time-averaged current $<j>_{av}$ on the initial
relative phases
$\phi_{12}(0)$ and $\phi_{23}(0)$ for (a) $\Lambda=$1, (b) 3 and (c) 10.}
\end{center}
\end{figure}

When the independent domains of the ordered phase  
grow up after the quench
of the system, as described in the introduction of this paper, 
the coupling constants between adjacent domains are
expected not to be constant but to increase with time.
Such time-dependent coupling constants change the motion
gradually from initially chaotic to periodic.
In a real system, every pair of condensates need not have
the same coupling constant $K_{ij}$.
When some dissipative term is introduced to eq. (1),
the calculation for $K_{12}>K_{23}=K_{31}$
shows that the first and second condensates are united, subsequently the united
condensate and the third one create the two-body motion; the dissipative term is
described in the following paper.

The above consideration is applicable to $n$ cyclically coupled BECs.
This system has $2\pi k/n$-phase modes about the stationary solutions
which result from the initial relative phases $\phi_{12}(0)=\phi_{23}(0)=
\cdots =\phi_{n-1,n}(0)=2\pi k/n$, where $k=0, 1, \cdots, n-1$.
In order to obtain the exact conclusions on CJC in $n$ BECs needs to carry 
out an analysis for
each value of $n$. However, the following estimation for weak coupling
roughly gives the dependence of CJC on $n$. Following the case of $n=3$,
we will define the ensemble-averaged CJC for $n$ BECs
\begin{eqnarray}
<j_{CJC}>_{n}=\frac{1}{(2\pi)^{n-1}}\int |<j>_{av}(\xv)|d\xv \\
\xv \equiv (\phi_{12}(0), \phi_{23}(0), \cdots, \phi_{n-1,n}(0)),
\end{eqnarray}
where
\begin{equation}
<j>_{av}(\xv)=\frac{1}{nT} \int_0^T \left( \sum_{k=1}^{n} j_{k\rightarrow k+1}
(t, \xv)\right) dt,
\end{equation}
is the time- and piece-averaged current starting from the initial phase $\xv$.
When the coupling is weak, as shown in Fig. 2(c), the main contribution
for CJC comes from the stationary solutions.
In the limit of weak coupling, we assume that only the stationary
solutions produce CJC:
\begin{equation}
<j_{CJC}>_{n}=\frac{1}{(2\pi)^{n-1}}\sum_{k=0}^{n-1} \frac{2}{n}
\left| \sin \left( \frac{2\pi k}{n} \right) \right|.
\end{equation}
It is easy to show that $<j_{CJC}>_{n}$ decreases with $n$ to disappear
for $n \rightarrow \infty$.
For example, we find $<j_{CJC}>_{n}=2.9 \times 10^{-2}, 4.0\times
10^{-3},
7.8\times 10^{-4}$ from eq. (17), respectively, for $n=3, 4, 5$.
Even if the coupling is not weak, the relative ratio $<j_{CJC}>_{n}/
<j_{CJC}>_{3}$ is not changed very much.
When independent domains are nucleated in a system, the number of $n$
cyclically
coupled BECs decreases with $n$.
Hence it is concluded that CJC arises mainly
from cyclic BECs with a small value of
$n$.
Following these results, we can estimate how the vortex density
in the KZ mechanism is reduced.
Then, the first assumption is that the vortices
are generated only from the CJC of $n=3$
cyclic
BECs based on the above considerations. The second is that the vortex density 
in the KZ mechanism is caused by
 the CJC due to the stationary solution
$\phi_{12}=\phi_{23}=2\pi/3$, i.e., $<j_{KZ}>_{3}=0.577$
which is the maximum of $<j_{CJC}>_{3}$,
indicating that CJC flows independently of the initial relative phases.
However, the CJC actually depends on the initial relative phases and
the coupling constant, so that $<j_{CJC}>_{3}$ is less
than $<j_{KZ}>_{3}$ to lessen
the density of generated vortices.
Figure 3 shows the dependence of $<j_{CJC}>_{3}/<j_{KZ}>_{3}$
on $\Lambda$, which can be related to the reduced ratio of the vortex
density.

\begin{figure}
\begin{center}
\epsfxsize=7cm \ \epsfbox{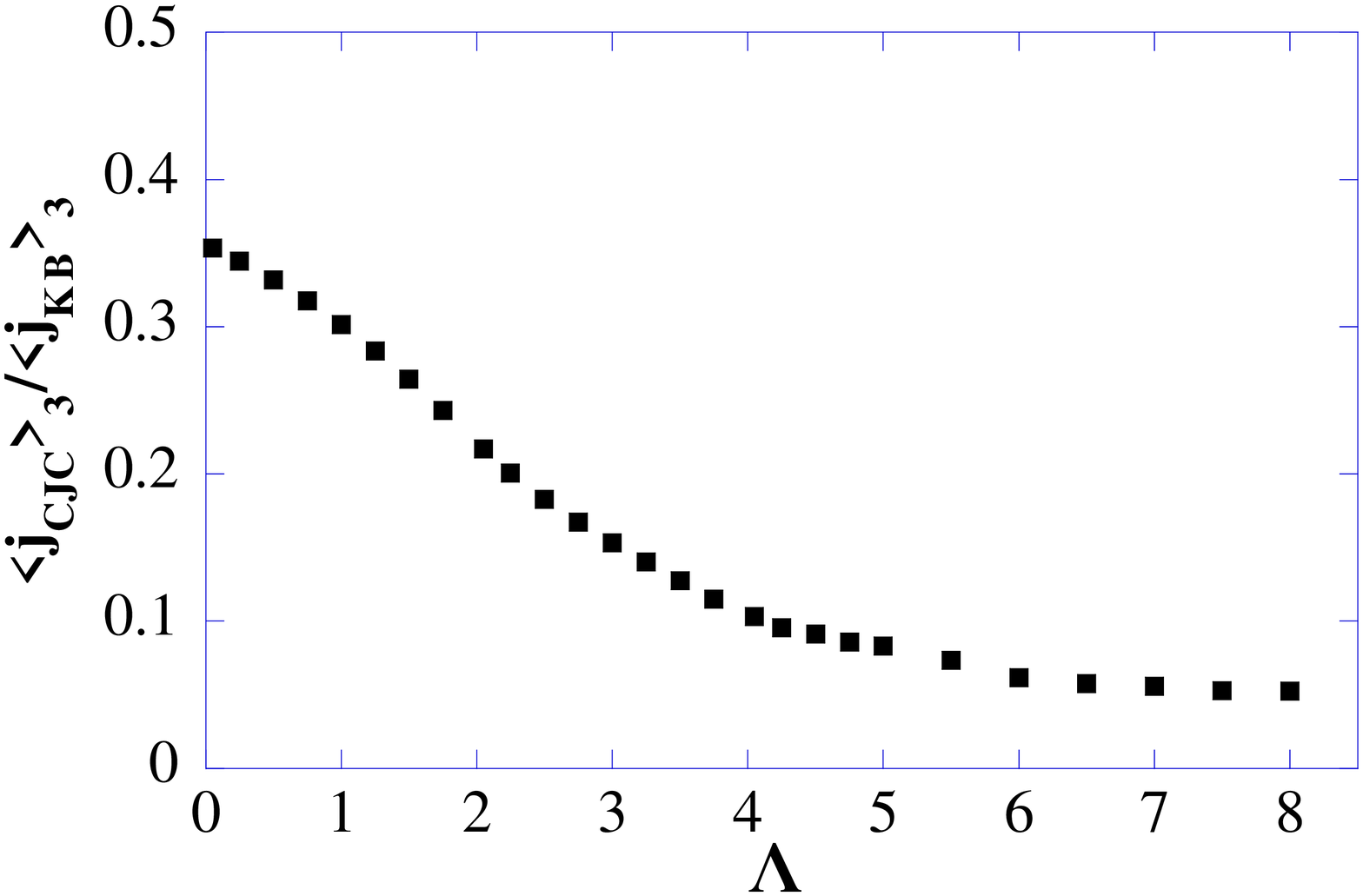} \\
\vspace{3mm}
\caption{Dependence of $<j_{CJC}>_{3}/<j_{KZ}>_{3}$
on $\Lambda$.}
\end{center}
\end{figure}

It is difficult to quantitatively estimate the coupling constants
in superfluid helium, although it may be possible in a dilute gas of
trapped atoms.
More detailed studies that include the application to a real system
and dissipative effects are to be published shortly.

\vskip 9pt
The authors thank T. Iida for useful discussions.

\end{document}